%% file: monopole.tex
\documentclass[12pt]{article}
\usepackage{epsfig}
\usepackage{rotating}
\usepackage{amsmath}
\usepackage{hhline}
\usepackage{amssymb}
\usepackage{times}
\usepackage{cite}

\newlength{\dinwidth}
\newlength{\dinmargin}
\begin{document}  
\newcommand{\pom}{{I\!\!P}}
\newcommand{\reg}{{I\!\!R}}
\newcommand{\slowpi}{\pi_{\mathit{slow}}}
\newcommand{\fiidiii}{F_2^{D(3)}}
\newcommand{\fiidiiiarg}{\fiidiii\,(\beta,\,Q^2,\,x)}
\newcommand{\n}{1.19\pm 0.06 (stat.) \pm0.07 (syst.)}
\newcommand{\nz}{1.30\pm 0.08 (stat.)^{+0.08}_{-0.14} (syst.)}
\newcommand{\fiidiiiful}{F_2^{D(4)}\,(\beta,\,Q^2,\,x,\,t)}
\newcommand{\fiipom}{\tilde F_2^D}
\newcommand{\ALPHA}{1.10\pm0.03 (stat.) \pm0.04 (syst.)}
\newcommand{\ALPHAZ}{1.15\pm0.04 (stat.)^{+0.04}_{-0.07} (syst.)}
\newcommand{\fiipomarg}{\fiipom\,(\beta,\,Q^2)}
\newcommand{\pomflux}{f_{\pom / p}}
\newcommand{\nxpom}{1.19\pm 0.06 (stat.) \pm0.07 (syst.)}
\newcommand {\gapprox}
   {\raisebox{-0.7ex}{$\stackrel {\textstyle>}{\sim}$}}
\newcommand {\lapprox}
   {\raisebox{-0.7ex}{$\stackrel {\textstyle<}{\sim}$}}
\def\gsim{\,\lower.25ex\hbox{$\scriptstyle\sim$}\kern-1.30ex%
\raise 0.55ex\hbox{$\scriptstyle >$}\,}
\def\lsim{\,\lower.25ex\hbox{$\scriptstyle\sim$}\kern-1.30ex%
\raise 0.55ex\hbox{$\scriptstyle <$}\,}
\newcommand{\pomfluxarg}{f_{\pom / p}\,(x_\pom)}
\newcommand{\dsf}{\mbox{$F_2^{D(3)}$}}
\newcommand{\dsfva}{\mbox{$F_2^{D(3)}(\beta,Q^2,x_{I\!\!P})$}}
\newcommand{\dsfvb}{\mbox{$F_2^{D(3)}(\beta,Q^2,x)$}}
\newcommand{\dsfpom}{$F_2^{I\!\!P}$}
\newcommand{\gap}{\stackrel{>}{\sim}}
\newcommand{\lap}{\stackrel{<}{\sim}}
\newcommand{\fem}{$F_2^{em}$}
\newcommand{\tsnmp}{$\tilde{\sigma}_{NC}(e^{\mp})$}
\newcommand{\tsnm}{$\tilde{\sigma}_{NC}(e^-)$}
\newcommand{\tsnp}{$\tilde{\sigma}_{NC}(e^+)$}
\newcommand{\st}{$\star$}
\newcommand{\sst}{$\star \star$}
\newcommand{\ssst}{$\star \star \star$}
\newcommand{\sssst}{$\star \star \star \star$}
\newcommand{\tw}{\theta_W}
\newcommand{\sw}{\sin{\theta_W}}
\newcommand{\cw}{\cos{\theta_W}}
\newcommand{\sww}{\sin^2{\theta_W}}
\newcommand{\cww}{\cos^2{\theta_W}}
\newcommand{\trm}{m_{\perp}}
\newcommand{\trp}{p_{\perp}}
\newcommand{\trmm}{m_{\perp}^2}
\newcommand{\trpp}{p_{\perp}^2}
\newcommand{\alp}{\alpha_s}

\newcommand{\alps}{\alpha_s}
\newcommand{\sqrts}{$\sqrt{s}$}
\newcommand{\LO}{$O(\alpha_s^0)$}
\newcommand{\Oa}{$O(\alpha_s)$}
\newcommand{\Oaa}{$O(\alpha_s^2)$}
\newcommand{\PT}{p_{\perp}}
\newcommand{\JPSI}{J/\psi}
\newcommand{\sh}{\hat{s}}
\newcommand{\uh}{\hat{u}}
\newcommand{\MP}{m_{J/\psi}}
\newcommand{\PO}{I\!\!P}
\newcommand{\xbj}{x}
\newcommand{\xpom}{x_{\PO}}
\newcommand{\ttbs}{\char'134}
\newcommand{\xpomlo}{3\times10^{-4}}  
\newcommand{\xpomup}{0.05}  
\newcommand{\dgr}{^\circ}
\newcommand{\pbarnt}{\,\mbox{{\rm pb$^{-1}$}}}
\newcommand{\gev}{\,\mbox{GeV}}
\newcommand{\WBoson}{\mbox{$W$}}
\newcommand{\fbarn}{\,\mbox{{\rm fb}}}
\newcommand{\fbarnt}{\,\mbox{{\rm fb$^{-1}$}}}

\newcommand{\meter}{\mbox{\rm ~m}}
%
%
\newcommand{\qsq}{\ensuremath{Q^2} }
\newcommand{\gevsq}{\ensuremath{\mathrm{GeV}^2} }
\newcommand{\et}{\ensuremath{E_t^*} }
\newcommand{\rap}{\ensuremath{\eta^*} }
\newcommand{\gp}{\ensuremath{\gamma^*}p }
\newcommand{\dsiget}{\ensuremath{{\rm d}\sigma_{ep}/{\rm d}E_t^*} }
\newcommand{\dsigrap}{\ensuremath{{\rm d}\sigma_{ep}/{\rm d}\eta^*} }
\def\Journal#1#2#3#4{{#1} {\bf #2} (#3) #4}
\def\NCA{Nuovo Cimento}
\def\RPP{Rep. Prog. Phys.}
\def\ARNPS{Ann. Rev. Nucl. Part. Sci.}
\def\NIM{Nucl. Instrum. Methods}
\def\NIMA{{Nucl. Instrum. Methods} {\bf A}}
\def\NPB{{Nucl. Phys.}   {\bf B}}
\def\NPPS{Nucl. Phys. Proc. Suppl.} 
\def\NPPSC{{Nucl. Phys. Proc. Suppl.} {\bf C}}
\def\PR{Phys. Rev.}
\def\PLB{{Phys. Lett.}   {\bf B}}
\def\PRL{Phys. Rev. Lett.}
\def\PRD{{Phys. Rev.}    {\bf D}}
\def\PRC{{Phys. Rev.}    {\bf C}}
\def\ZPC{{Z. Phys.}      {\bf C}}
\def\EJC{{Eur. Phys. J.} {\bf C}}
\def\EPL{{Eur. Phys. Lett.} {\bf}}
\def\CPC{Comp. Phys. Commun.}
\def\NP{{Nucl. Phys.}}
\def\JPG{{J. Phys.} {\bf G}} 
\def\EPC{{Eur. Phys. J.} {\bf C}}
\def\PRSL{{Proc. Roy. Soc.}} {\bf}
\def\PETF{{Pi'sma. Eksp. Teor. Fiz.}} {\bf}
\def\JETPL{{JETP Lett}}{\bf}
\def\IJTP{Int. J. Theor. Phys.}
\def\HJ{Hadronic J.}


\begin{titlepage}

\begin{flushleft}
DESY 04-240 \hfill ISSN 0418-9833 \\
December 2004
\end{flushleft}

\vspace{2cm}

\begin{center}
\begin{Large}
 
{\boldmath \bf A Direct Search for Stable Magnetic Monopoles Produced in 
Positron-Proton Collisions at HERA} \\
\vspace*{2.0cm} H1 Collaboration \\
\end{Large}
\end{center}

\vspace*{2.5cm}

\begin{abstract}
\noindent

A direct search has been made for magnetic monopoles produced in 
$e^+p$ collisions at a centre of mass energy of 300 GeV at HERA.  
The beam pipe surrounding the interaction region in 1995-1997 
was investigated using a SQUID magnetometer to look for stopped 
magnetic monopoles. During this time an integrated luminosity of  
62 pb$^{-1}$ was delivered. No magnetic monopoles were observed and 
charge and mass dependent upper limits on the $e^+p$ production 
cross section are set.

\end{abstract}
\begin{center}
(Submitted to European Physical Journal C)
\end{center}

\end{titlepage}

\begin{flushleft}
  \input{h1auts}
\end{flushleft}

\newpage

\section{Introduction}
\label{Vacuo}

One of the outstanding issues in modern physics is the question of the  
existence of magnetic monopoles. Dirac showed that their existence  
leads naturally to an explanation of electric charge 
quantisation\cite{dirac}. Magnetic monopoles 
are also predicted from field theories which unify the fundamental 
forces\cite{gut1,gut2,mtheory1,mtheory2}. Furthermore,  the formation of 
a monopole condensate provides a possible mechanism for quark 
confinement\cite{confine}.  Nevertheless, despite a large number of 
searches\cite{searchgen,search2} using a variety of experimental 
techniques, no reproducible evidence has been found to support the 
existence of  monopoles. Searches for magnetic monopoles produced in 
high energy particle collisions have been made in  
$p\bar{p}$\cite{D01,D02,ppd0} and $e^+e^-$\cite{ee1,ee2,ee3,ee4,ee5,eegam}
interactions. This paper describes the first search for monopoles produced 
in high energy $e^+p$ collisions.  

The quantisation of the angular momentum of a system of an electron with 
electric charge $e$ and a monopole with magnetic charge $g$ leads to 
Dirac's celebrated charge quantisation condition $eg=n \hslash c/2$, where 
$\hslash$ is Planck's constant divided by $2\pi$, $c$ is the speed of 
light and $n$ is an 
integer\cite{dirac}. Within this approach, taking $n=1$ sets the 
theoretical minimum magnetic charge which can be possessed by a particle 
(known as the Dirac magnetic charge, $g_D$). However, if the elementary 
electric charge is considered to be held by the down quark then the minimum 
value of this fundamental magnetic charge will be three times larger. 
The value of the fundamental magnetic charge could be 
even higher since the application of the Dirac argument to a particle 
possessing both electric and magnetic charge (a so-called 
dyon\cite{schwinger,othersw}) restricts the values of $n$ to be 
even\cite{schwinger}.

Monopoles are also features of current unification theories such as 
string theory \cite{mtheory1,mtheory2} and Supersymmetric Grand Unified 
Theories\cite{gut1,gut2}. Both of these approaches tend to predict 
heavy primordial monopoles with mass values in excess of $10^{15}$ 
GeV.  However, in some Grand Unified scenarios values of monopole mass 
as low as $10^4$ GeV\cite{so103,so101,so102} are allowed. Light 
monopoles are also predicted in other 
approaches\cite{other0,other1,other2,other3} and postulates on values of 
the classical radius of a monopole lead to estimates of mass of $\cal O$(10) 
GeV\cite{search2}.  

Since the value of the coupling constant of a photon to a monopole 
($\alpha_m \approx 34n^2$) is substantially larger than for a 
photon-electron interaction ($\alpha_e \approx 1/137$) perturbative 
field theory cannot be reliably used to calculate the rates of processes 
involving monopoles. 
The large coupling also implies that ionisation energy losses 
will be typically several orders of magnitude greater for monopoles than 
for minimum ionising electrically charged particles\cite{ahlen1,ahlen,ahlen2}.

Direct experimental searches using a variety of tracking devices to detect 
the passage of highly ionising particles with monopole properties have 
been made\cite{searchgen}. Direct searches have been made for
monopoles in cosmic rays \cite{cabrera} and for monopoles which stop in
matter such as at accelerators\cite{D02} and in lunar 
rock\cite{lunar1,lunar2,lunar3}. One method of detection is the search 
for the induction of a persistent current within a superconducting 
loop\cite{lunar1}, the approach adopted here.  
Measurements of multi-photon production in collider 
experiments \cite{ppd0,eegam} allow indirect searches for monopoles to 
be made, the interpretation of which is believed to be 
difficult \cite{dracula,kimc}.  


\section{The Experimental Method}

For the direct search reported here we use the fact that heavily ionising 
magnetic monopoles produced in $e^+p$ collisions may stop in the beam 
pipe surrounding the H1 interaction point at HERA. The binding energy of 
monopoles which stop in the material of the pipe (aluminium in the years 1995-1997) 
is expected to be large \cite{kimc1} and so they 
should remain permanently trapped provided that they are stable. The 
beam pipe was cut into long thin strips which were each passed through 
a superconducting coil coupled to a Superconducting Quantum 
Mechanical Interference Device (SQUID). Fig.~\ref{princ} shows a schematic 
diagram illustrating the principle of the method used.
Trapped magnetic monopoles in a strip will cause a 
persistent current to be induced in the superconducting coil 
by the magnetic field of the monopole, after complete passage of the strip 
through the coil.   
In contrast, the induced currents from the magnetic fields of the 
ubiquitous permanent magnetic dipole moments in the material, 
which can be pictured as a series of equal and opposite   
magnetic charges, cancel so that the current due to dipoles returns 
to zero after passage of the strip. 

The aluminium beam pipe used in 1995-1997 was exposed to a luminosity 
of $62 \pm 1\mathrm{pb}^{-1}$. 
The beam pipe around the interaction point had a
diameter of 9.0~cm and thickness 1.7~mm in the range 
$-0.3 < z < 0.5$~m \footnote{Here $z$ is the longitudinal coordinate
	defined with $z=0$ as the nominal positron-proton interaction point and
	with the positive $z$ axis along the proton beam direction.}
and a diameter of 11.0~cm and thickness 2~mm in the range $0.5 < z < 2.0$~m.
During HERA operations it was immersed in a 1.15 T solenoidal magnetic
field which was directed parallel to the beam pipe, along the $+z$
direction. 
This length of the pipe, covering $-0.3 < z < +2.0$~m,
was cut into 45 longitudinal strips each of length on average of 573~mm 
($\sim 2$~mm was lost at each cut).  
The central region ($-0.3 < z < 0.3$~m) was cut into 15 long strips of 
width $\sim 18$~mm, two of which were further divided into 32 short 
segments varying in length from 
1 to 10~cm. The downstream region ($0.3 < z < 2.0$~m) was divided into 
3 longitudinal sections each of which was cut into 10 long strips of width 
$\sim 32$~mm. 
The long strips and short segments were each passed along the 
axis of the 2G Enterprises type 760 magnetometer 
\cite{2G} at the Southampton Oceanography Centre, UK. This is a warm bore 
device with high sensitivity and low noise level. The samples of strips 
and segments were passed 
through the magnetometer in steps, pausing after each step, after which 
the current in the superconducting loop was measured. The residual 
persistent current after the complete traversal of a sample through the 
loop was measured by taking the difference in the measured current after 
and before passage.  The readings for each sample were repeated 
several times. This allowed the reproducibility of the results to be 
studied so that random flux jumps and base line drifts could be 
identified. Any real monopole trapped in the pipe would give a consistent 
and reproducible current step.  

A long, thin solenoid, wound with copper wire on a cylindrical copper former, 
was used to assess the sensitivity of the SQUID magnetometer to a monopole. 
The magnetic field outside of the ends of a long solenoid is similar to 
that produced by a monopole. A solenoid can thus be considered as 
possessing two oppositely charged ``pseudopoles'' of pole strength 
$g = N \cdot I \cdot S/g_D$ in units of the Dirac magnetic charge. Here 
$N$ is the number of turns per metre length, $I$ is the current and 
$S$ is the cross sectional area of the coil and $g_D = 3.3 \times 10^{-9}$ 
Amp\`ere-metres is the Dirac magnetic charge introduced 
above. Hence the current and radius of the solenoid can be chosen to 
mimic the desired pole strength.\footnote{A numerical study integrating the 
Biot-Savart equation for the magnetic field outside the dimensions of the 
magnetometer coil showed that this simple formula is accurate to
better than $\pm 3\%$.} 
To calibrate, the solenoid was stepped through the magnetometer. Data 
were taken with different currents subtracting the measurements with zero 
current to correct for the dipole impurities in the coil and its former.  
The measured increase in current in the magnetometer following the 
passage of one end of the solenoid was found to vary linearly with the 
pseudopole strength.   
Four coils, the details of which are given in table~\ref{coils}, were used 
at different times for the calibration procedure.  
Fig.~\ref{calib1} shows the magnetometer current divided by the solenoid 
pole strength as a function of the 
pole strength in units of $g_D$. There is a good consistency 
between the calibrations and the magnetometer is linear over more 
than 2 orders of magnitude in pole strength. The uncertainty on 
the point at the lowest pole strength is large because the current step 
size is small at such a low excitation of the calibration coil. Small  
differences, at the level of $\sim 0.03 g_D$, of the current readings 
occurred between traversals, presumably due to the system picking up 
specks of slightly magnetised dust between the traversals. Such differences 
show up as noise for low excitations but are less important at higher 
excitation of the calibration coil. The mean values of the 
calibration factor, together with the 
root mean square deviations, from each coil are given in table~\ref{coils}. 

To simulate trapped monopole behaviour the long solenoid was placed 
along a beam pipe strip and both were passed (jointly) through the 
magnetometer. Only one end of the calibration solenoid was allowed 
to traverse the magnetometer hence simulating the passage of a monopole 
in the strip. Fig.~\ref{dummy} shows the absolute value of the measured 
magnetometer current as the strip alone was stepped through and when 
pseudopoles of values $2.3g_D$ and $-2.3g_D$ were attached to the strip.   
The large structure at the centre comes from the magnetic fields from 
the permanent magnetic dipole moments in the aluminium. The final current 
persists when the pseudopoles are present. When the pseudopoles are absent 
the current returns to zero despite the very large permanent dipole moments. 
The inset of Fig.~\ref{dummy} shows the value of the measured current 
(on a linear scale) as the strip leaves the magnetometer coil. The values 
of current for these pseudopole strengths are equal and opposite and at 
the value expected from the calibration performed purely with the solenoid. 
This proves that a magnetic monopole attached to the beam 
pipe would have been detected by the magnetometer. 
 
\section{Results}

\subsection{Magnetometer Scans}

The data were taken in four separate sets, taking just over one day   
for each set: in December 2002, January 2003, May 2003 and 
January 2004.  In the first two sets of data (Dec. 2002 and Jan. 2003) 
all the strips from the central beam pipe, covering $-0.3 < z < 0.3$ m, 
were passed through the magnetometer once for each measurement.     
The values of the residual persistent current were computed from  
the difference between the first reading, typically $\sim 20$ cm before 
the strip entered the magnetometer, and the last reading which came 
typically $\sim 30$ cm after the strip left the magnetometer. These were 
then converted to Dirac Monopole units ($g_D$) by dividing by the 
calibration constant, determined as described above. The results for the 
long strips are shown in Fig.~\ref{overview_central}.  
In the first dataset (Dec. 2002) only single measurements were made on 
each long strip (except strip 13) and these are 
shown as open circles in Fig.~\ref{overview_central}. Two of the strips 
measured showed 
persistent currents of value expected from the passage of a magnetic charge 
of about $+0.7 g_D$. Here a positive pole is defined to be a North seeking
pole, i.e. one that is accelerated in the $+z$ direction by the H1 magnetic 
field. All the strips were then remeasured several times 
in the second set of data (Jan. 2003), shown as closed circles in 
Fig.~\ref{overview_central}. None of them (except a single reading for 
strip 3) showed a persistent current 
after traversal through the magnetometer. It was therefore assumed that the 
observed persistent currents during the first set had been caused by 
random jumps in the base line of the magnetometer electronics.  
It can be seen from Fig.~\ref{overview_central} that none of the strips
showed a persistent current which appeared consistently in more than one 
reading. Fig.~\ref{overview_small} shows the results of the 
measurements on the 32 short segments from the central beam pipe. 
Sample 6 showed a reading at $\sim0.5 g_D$ for the first measurement  
but the measurement was compatible with zero when it was remeasured.   
It was therefore assumed that the first reading was due to a base line 
shift. We concluded that  
none of the long strips or short segments showed any consistent signal for 
a monopole in multiple readings and therefore that there were 
no magnetic monopoles trapped in the central section of the beam pipe. 
 
In the third and fourth sets of data (May 2003, Jan. 2004) the strips from 
the downstream beam pipe were investigated. In these sets of data the 
magnetometer measurements proved to be less stable. This was probably due 
to the induced currents from the large permanent dipole moments encountered 
which caused the magnetometer to lose its memory of the zero 
level.\footnote{The cause of this was not understood.}  
The permanent dipole moments in the downstream beam pipe were found to be  
much larger than those in the central section.  These dipole moments were 
all observed to be aligned along the H1 magnetic 
field, i.e. in the same direction as the proton beam.  
In these datasets the strips were passed through twice, first with the 
strip length parallel to the proton beam, termed $+z$ end first, 
and then with the strip aligned in the opposite direction, i.e. the 
length antiparallel to the proton beam direction, termed $-z$ end first. 
Fluctuations of the base level of size of $0.7 g_D$ were observed 
to happen much more frequently than for the central beam pipe section 
(Dec. 2002 and Jan. 2003 data) and in a more systematic way. 
Fluctuations of $-0.7 g_D$ were 
consistently present for the traversal with the $-z$ end first and 
fluctuations of $+0.7 g_D$ were observed in the first $25\%$ of the   
traversals with the $+z$ end first but absent in the remaining $75\%$ 
of the readings in this orientation. To check that a 
monopole would have been seen, even in these adverse conditions, the 
long calibration coil was placed on a strip and the procedure repeated. 
The expected deviations from the calibration coil were seen superimposed 
on the base line shifts, confirming that a trapped monopole would have 
been seen had it existed, even in these conditions. 

The permanent dipole moments in the downstream strips 
gave readings in the magnetometer up to 3 orders of magnitude larger 
than the final persistent current expected from a Dirac monopole. 
This is larger than those seen in Fig.~\ref{dummy}. It was found that 
the base line shifts could be avoided by demagnetising the strips in a 
low frequency decreasing magnetic field of initial strength 0.1 T. 
This is less than $10\%$ of the H1 magnetic field.  
The binding energy of monopoles in aluminium, the main constituent of the 
beam pipe, is thought to be hundreds of keV \cite{kimc1} compared to 
those of atoms which are at the eV level. Hence it was thought that such 
magnetic fields would be unable to dislodge a trapped monopole. 
All the strips were therefore subject to such a demagnetising field and 
remeasured.    
Demagnetisation was found to reduce the permanent dipole moments in the 
aluminium strips by about a factor of 20. After this procedure no further 
base line shifts were observed after passage of a strip.    

Fig.~\ref{overview} shows all the readings of persistent current    
after demagnetisation plotted against sample number. 
The strips were passed through several at a time for these data. 
Sample 17 consists of the 13 strips of the central 
beam pipe (shown individually in Fig.~\ref{overview_central}) passed 
through the magnetometer as a bundle. It is concluded    
from Figs.~\ref{overview_small} and \ref{overview}, that no monopole of 
strength greater than 
0.1 Dirac magnetic charge unit had stopped in any of the measured pieces  
which constituted $93 \pm 3 \%$ of the total beam pipe. The remainder  
was lost in the cutting procedure.        

\subsection{Upper Limits on the measured Cross Sections}

To derive an upper limit on the measured cross section it is necessary 
to compute the acceptance, i.e. the fraction of the monopoles produced 
which would have been detected. A model of the production process is 
therefore needed. Two models were used to compute the acceptance 
by Monte Carlo technique. In each of these a 
monopole - antimonopole ($M \bar M$) pair was assumed to be produced by 
a photon-photon interaction. The first model (model A) assumed spin 0 
monopole pair production by the elastic process 
$e^+ p \rightarrow e^+ M \bar M p $  
through the interactions of a photon radiated from each of the electron 
and proton. The proton was assumed to have the simple dipole form factor 
$1/(1+Q^2/0.71\gevsq)$, where $Q^2$ is the negative square of the four momentum 
transferred to 
the proton. The second model (model B) assumed spin 1/2 monopole pair 
production by the inelastic process 
$e^+ p \rightarrow e^+ M \bar M X $ (where $X$ is any state) through 
a photon-photon fusion interaction with a photon radiated from the electron 
and one radiated from a quark in the proton. The photon is radiated  
with a simple distribution given by $(1-\eta)^5/\eta$, with 
$\eta$ the fraction of the proton's energy carried by the photon. 
While the models implement the kinematic correlations in each event 
it should be noted that they depend on perturbation theory and 
therefore the predicted cross sections are unreliable, as mentioned 
previously. Events were generated 
according to model A using the programme CompHEP \cite{CompHep} and 
using a dedicated programme for model B. The generated final state particles 
were tracked through the H1 magnetic field to the beam pipe. 
If the thickness of beam pipe traversed was greater than the calculated 
range of the monopole in aluminium, it was assumed to stop.   
In this way the fraction of monopoles, which were detected by stopping 
in the beam pipe, was computed.  

Monopoles experience a force $g{\bf B}$ in a magnetic field ${\bf B}$. 
With the field aligned along the $z$ axis they have a parabolic 
trajectory with 
\begin{equation}
z(r)-z_v = 0.5 \frac{g|{\bf B}| r^2}{e P_T \beta_T} + \frac{r}{\tan \theta_0} 
\end{equation}
where $z_v$ is the $z$ coordinate of the vertex and $z(r)$ 
is the coordinate of a point on the trajectory at distance $r$ from the 
proton beam. The transverse momentum and 
tranverse velocity of the monopole are $P_T$ and $\beta_T$, respectively. 
The initial angle of the monopole to the proton beam direction is $\theta_0$
and $e$ is the unit of electric charge.  In this equation $g$ is the 
magnetic pole strength which is negative (positive) for South (North) poles 
which decelerate (accelerate) in the $+z$ direction in the H1 magnetic field.
The geometric acceptance is the fraction of the monopoles which traverse 
the beam pipe in the sampled length. The total acceptance is this fraction 
times the fraction which stop in the pipe. The range of monopoles in 
aluminium was computed by integrating the stopping power, $dE/dx$, given 
in \cite{ahlen} adjusted for the electron density in aluminium. 
Fig.~\ref{RErel} shows the computed range (normalised to mass), for
monopoles of strength 
$g_D$, versus $P/M=\beta \gamma$ where $P$ and $M$ are the momentum and 
mass of the monopole, respectively, and $\beta,\gamma$ are its velocity 
factors. The stopping power was computed in \cite{ahlen1,ahlen,ahlen2} 
by classically considering the long range monopole interactions with 
atomic electrons and hence the result is unaffected by the limitations 
of the use of perturbation theory.     

Figs.~\ref{DISaccB} and \ref{DISaccF} show the total efficiency for 
stopping a monopole in the beampipe for models A and B,  
respectively. These were computed for magnetic charges of 
$1$, $2$, $3$ and $6 g_D$, using the range calculations shown in Fig.~\ref{RErel},   
divided by the square of the monopole charge considered. The 
choice of magnetic charges was motivated by the Dirac quantisation 
condition \cite{dirac} or the Schwinger modification \cite{schwinger} 
applied to the electron as the fundamental unit of electric charge 
(magnetic charges = $1 g_D$ and $2 g_D$) or to the $d$ quark as 
the fundamental unit of electric charge (magnetic charges = $3 g_D$ 
and $6 g_D$). The acceptance increases rapidly as the magnetic charge 
increases since larger charges have higher $dE/dx$ so that a greater 
fraction of the monopoles stop in the beam 
pipe. Hence the curve for $6 g_D$ will also be 
approximately the acceptance for higher charged monopoles. 
The acceptance for South poles is somewhat larger than that for North 
poles since they decelerate in the H1 magnetic field, losing some energy, 
so that they stop more readily in the beam pipe.   
Higher mass monopole pairs are produced at smaller angles 
to the proton beam and tend to hit the downstream beam pipe. However, they 
are more energetic than for lower masses. Hence high masses with low magnetic 
charge pass through the downstream pipe whereas higher magnetic charges stop. 
This accounts for the rise in the acceptance at higher masses for  
magnetic charges of $2g_D$ and $3g_D$. The efficiencies for model A 
tend to be smaller than 
those for model B since in the latter the monopoles have a smaller 
mean transverse momentum than in the former which leads to a 
greater fraction of monopoles stopping in the beam pipe.   

The upper limit on the cross section for monopole-antimonopole pair 
production was derived within the context of each model, as follows. 
The failure to observe a monopole candidate means that there is an 
upper limit of 3 monopole pair events produced at the  
95$\%$ confidence level.
%
The cross section upper limit is then calculated from this, taking 
into account the uncertainties in the measured integrated luminosity,
in the fraction of the pipe surviving the cutting procedure, and the
statistical uncertainty in the acceptance computed from the models 
described above.
Here the acceptance is the fraction of the monopole pairs which produce either one 
or both monopoles which stop in the beam pipe. 
Fig.~\ref{ltBoson} shows the upper limit on the cross section 
at 95$\%$ confidence level for monopoles of strength $1$, $2$, $3$ and $6 g_D$ 
using acceptances determined from model A. Fig.~\ref{ltFermion} 
shows the upper limits determined using the acceptances from model B. 

Several other experiments have also produced limits on monopole production
cross sections for different masses and
charges\cite{D01,D02,ee1,ee2,ee3,ee4,ee5,lunar3}.
However, owing to the lack of a reliable field theory for monopole
production, different model assumptions were made in their derivations.  
Furthermore, although a universal production mechanism for monopole
production can be postulated, comparisons of cross section limits in
processes as diverse as $e^+p$, $p\bar{p}$ and $e^+e^-$ are difficult. 
This is the first search in $e^+p$ collisions. It could provide a 
sensitive testing ground for magnetic monopoles 
if a monopole condensate is responsible for quark confinement \cite{confine}. 
 
\section{Conclusions}

A direct search for magnetic monopoles produced in $e^+p$ collisions
at HERA at a centre of mass energy of $\sqrt s=300$ GeV has been made for 
the first time. Monopoles trapped in the beam pipe surrounding the 
interaction point were sought using a SQUID magnetometer which was 
sensitive down to 0.1 Dirac magnetic charges ($0.1g_D$). No monopole signal 
was observed.   
Upper limits on the monopole pair production cross section have been set for 
monopoles with magnetic charges from $1$ to $6 g_D$ or more and up to a 
mass of 140 GeV within the context of the models described. 

\section{Acknowledgement}
We are grateful to the HERA machine group whose outstanding efforts 
have made this experiment possible. We thank the engineers and technicians 
for their work in constructing and now maintaining the H1 detector, our 
funding agencies for financial support, the DESY technical staff for 
continual assistance and the DESY directorate for support and for the 
hospitality which they extend to non-DESY members of the collaboration. 
We thank Andrew Robertson of the Southampton Oceanography Centre for 
making the SQUID magnetometer available to us and for his assistance and 
hospitality. We are indebted to Barbara Maher 
and Vassil Karloukovski of the University of Lancaster Geography 
Department and Gordon Donaldson and Colin Pegrum of the University of 
Strathclyde Physics Department for valuable discussions and assistance 
during the early part of the experiment.


\begin{table}[h]
\begin{center}
\caption{ Description of the calibration coils.}
\vspace*{2mm}
\begin{tabular}{|c|c|c|c|c|}
\hline
 Coil & 1 & 2 & 3 & 4  \\
\hline
\hline
Core diam.(mm) & 6.08 & 3.15 & 2.1 & 2.1 \\
\hline
Coil length (mm) & 700 & 700 & 300 & 300 \\
\hline
Wire diam.(mm) (including insulation) & 0.18 & 0.18 & 0.1 & 0.1 \\
\hline
Turns per metre & 11000 & 11000 & 10000 & 30000 \\
\hline
Coil area ($S$ mm$^2$) (including wire) & 32.6 & 9.7 & 3.80 & 4.54 \\
\hline
Uncertainty in area & 3.3$\%$ & 6.0$\%$ & 10$\%$ & 10$\%$ \\
\hline 
Mean magnetometer current per $g_D$ & $1.10$ & $1.09$ & $1.11$ & $1.11$ \\
(arbitrary units) & & & & \\
\hline
R.M.S. deviation of the readings & 0.016 & 0.024 & 0.030 & 0.017 \\
\hline 
\end{tabular}
\label{coils}
\end{center}
\end{table}

\input{plots}

\end{document}

%% file: h1auts.tex

A.~Aktas$^{10}$,               
V.~Andreev$^{26}$,             
T.~Anthonis$^{4}$,             
S.~Aplin$^{10}$,               
A.~Asmone$^{34}$,              
A.~Babaev$^{25}$,              
S.~Backovic$^{31}$,            
J.~B\"ahr$^{39}$,              
A.~Baghdasaryan$^{38}$,        
P.~Baranov$^{26}$
% ...

\bigskip{\it
% ...
$ ^{4}$ Inter-University Institute for High Energies ULB-VUB, Brussels;
          Universiteit Antwerpen, Antwerpen; Belgium$^{ c}$ \\
% ...
$ ^{10}$ DESY, Hamburg, Germany \\
% ...
$ ^{25}$ Institute for Theoretical and Experimental Physics,
          Moscow, Russia$^{ l}$ \\
 $ ^{26}$ Lebedev Physical Institute, Moscow, Russia$^{ e}$ \\
% ...
$ ^{31}$ Faculty of Science, University of Montenegro,
          Podgorica, Serbia and Montenegro \\
% ...
$ ^{34}$ Dipartimento di Fisica Universit\`a di Roma Tre
          and INFN Roma~3, Roma, Italy \\
% ...
$ ^{38}$ Yerevan Physics Institute, Yerevan, Armenia \\
 $ ^{39}$ DESY, Zeuthen, Germany \\
% ...
$ ^c$ Supported by FNRS-FWO-Vlaanderen, IISN-IIKW and IWT
      and  by Interuniversity
Attraction Poles Programme,
      Belgian Science Policy \\
% ...
$ ^e$ Supported by the Deutsche Forschungsgemeinschaft \\
% ...
$ ^l$ Partially Supported by Russian Foundation
      for Basic Research, grant    no. 00-15-96584 \\
}

%% file: plots.tex

\begin{figure}
\begin{center}
 \epsfig{file=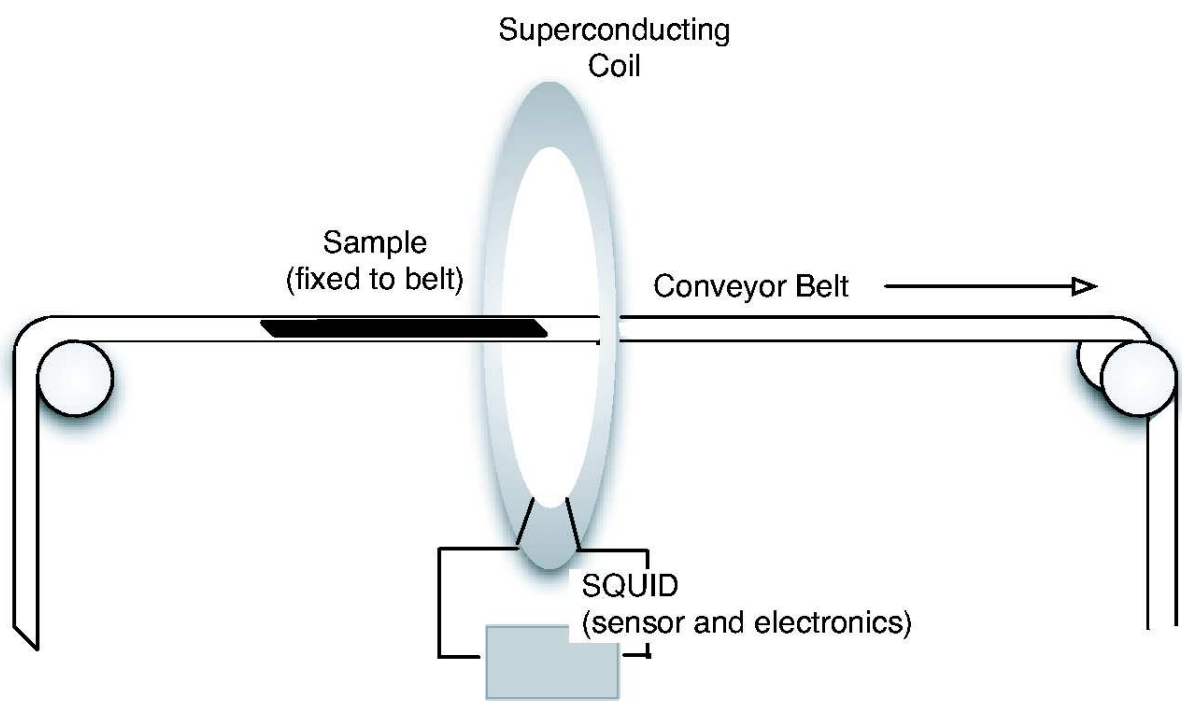,width=16cm}
\caption{The schematic diagram shows the principle of the method. The 
conveyor belt travelled in steps of typically 5 cm until the sample 
traversed completely the superconducting coil. At each step the conveyor 
belt stopped for $1$ sec before the current in the 
superconducting coil (magnetometer current) was read to avoid the effects of eddy currents. 
The time for each step was typically $3$ secs.}
\label{princ}
\end{center}
\vspace*{3cm}
\end{figure}

\begin{figure}
\begin{center}
\vspace{-20mm}
\epsfig{file=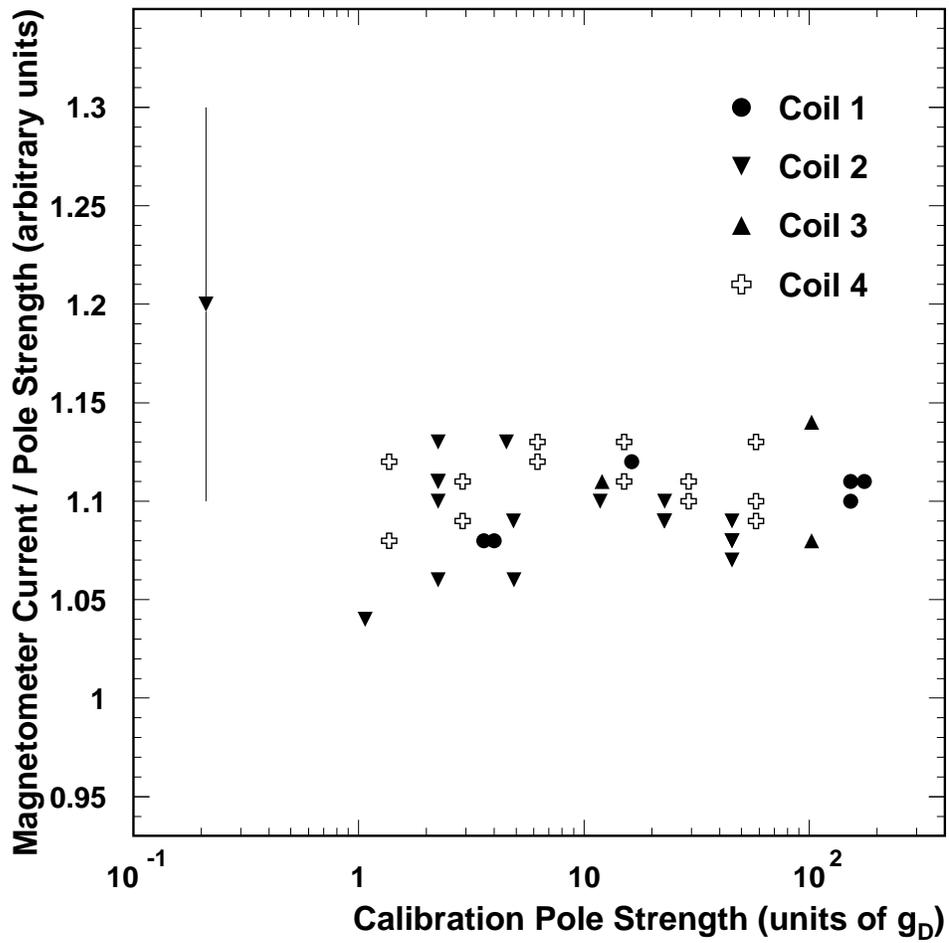,width=14.2cm}
\caption{The magnetometer current divided by the calibration pole strength  
as a function of the calibration pole strength for the four coils used 
(see table \ref{coils} for details of the coils).}
\label{calib1} 
\end{center}  
\end{figure}

\begin{figure}
\hspace{-20mm}
\epsfig{file=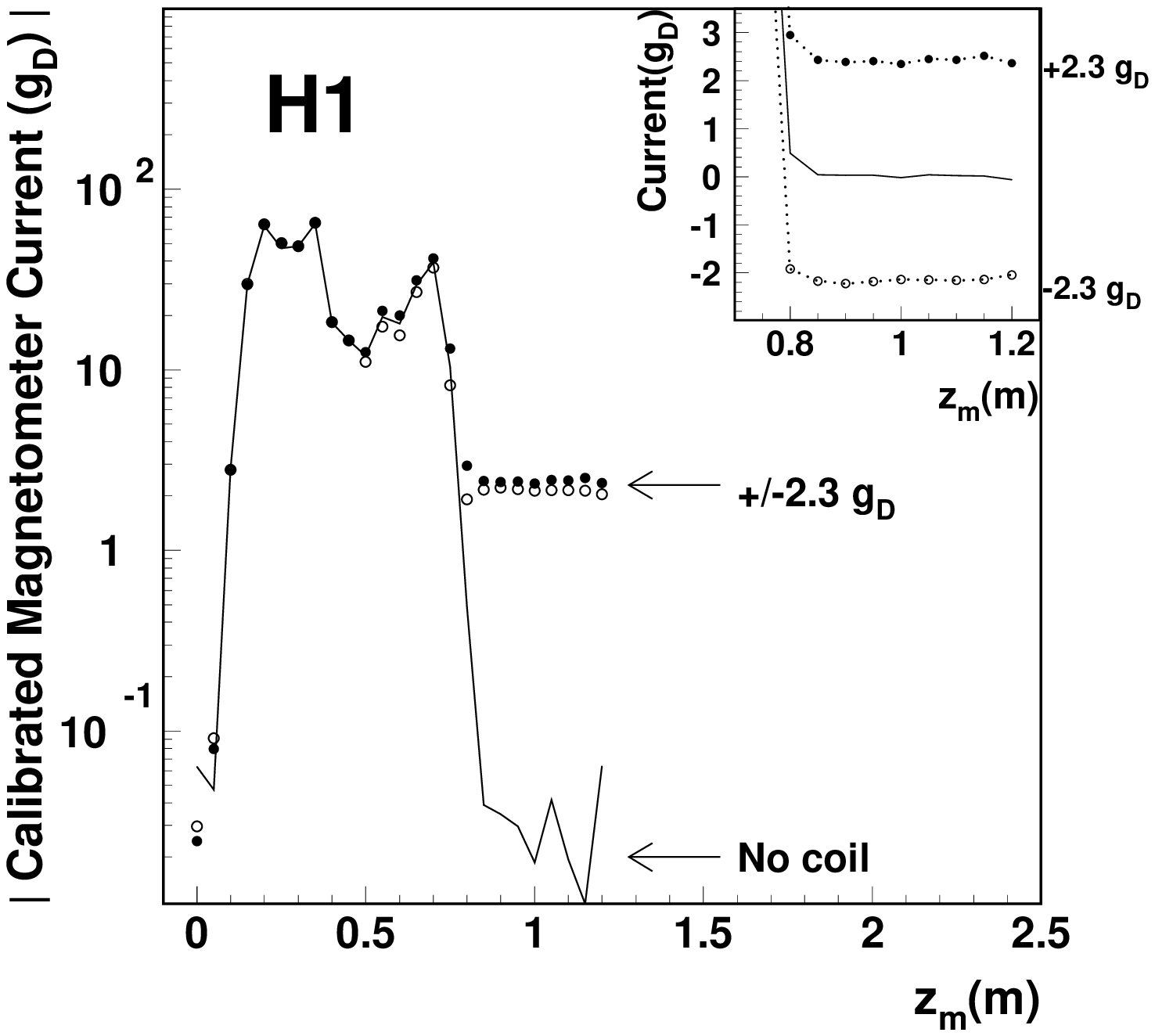,
width=21cm}
\vspace*{-14cm}
\caption[junk]{ The absolute (unsigned) value of the calibrated magnetometer current 
on a logarithmic scale versus step position ($z_m$) for 
a strip from the central beam pipe region ($-0.3 < z < 0.3 \meter$). The solid 
line shows the measurements with the long strip alone. The closed (open) 
points show the measurements with the long strip together with the 
calibration solenoid excited to simulate a 
pole of strength $+2.3g_D$ ($-2.3g_D$). The inset shows the signed 
measurements of the calibrated magnetometer currents versus the step position for 
$z_m > 0.8$ m on a linear scale. The expected persistent currents for 
monopoles of strength $\pm 2.3 g_D$ are shown by the arrow on the 
logarithmic plot and by the numbers in the margin on the inset linear plot.}
\label{dummy}
\end{figure}

\begin{figure}
\begin{center}
\vspace{-20mm}
\epsfig{file=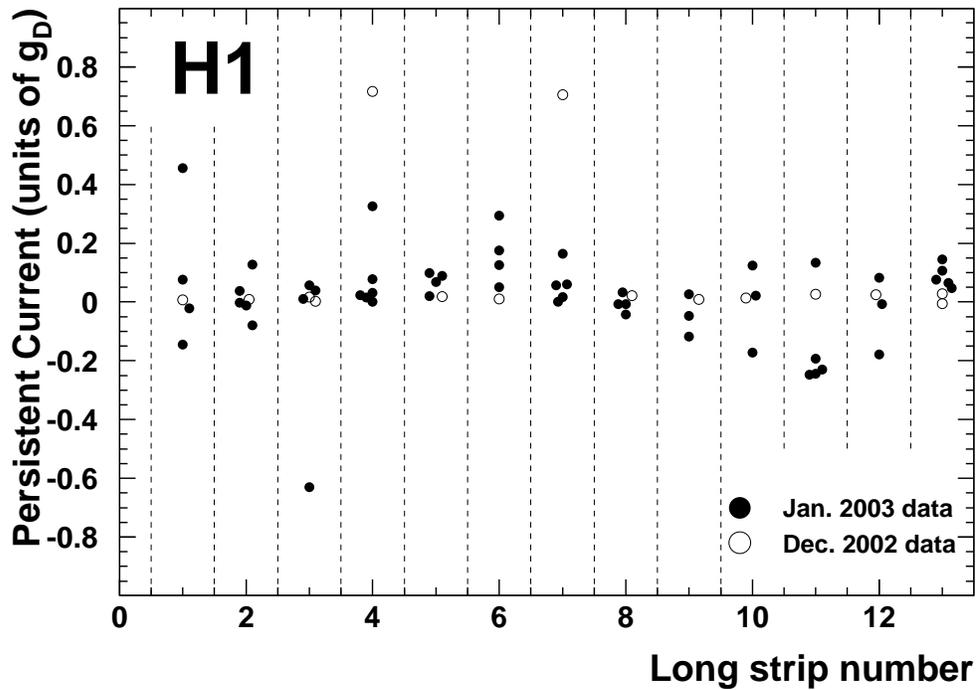,width=14.2cm}
\caption{The measured persistent current (in units of $g_D$) in each strip, 
after passage 
through the magnetometer, plotted against 
strip number for 13 strips of the central beam pipe. Some of the strip 
numbers are offset for clarity. It can be seen that none of the 
fluctuations observed in single readings occured consistently in other 
readings on the same strip showing that no trapped monopole was 
present.}
\label{overview_central}
\end{center}
\end{figure}

\begin{figure}
\begin{center}
\vspace{-20mm}
\epsfig{file=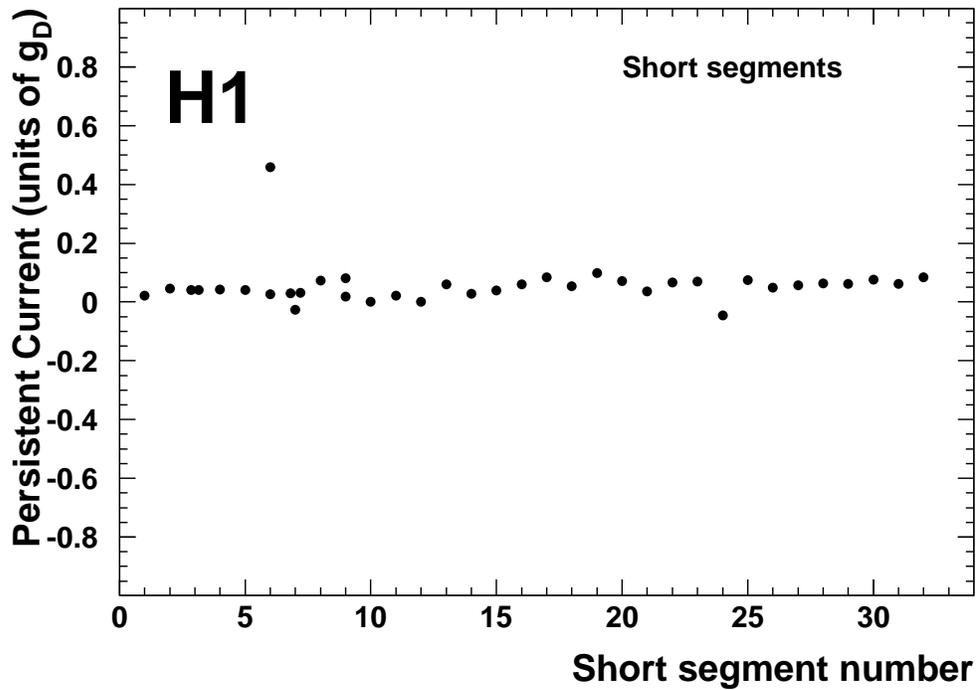,width=14.2cm}
\caption{The measured persistent currents (in units of $g_D$), after 
passage through the magnetometer, plotted against sample number for the 
two strips of the central beam pipe 
which were cut into short segments. It can be seen that none of the 
fluctuations observed in single readings occured consistently in other 
readings on the same sample showing that no trapped monopole was 
present.}
\label{overview_small}
\end{center}
\end{figure}

\begin{figure}
\begin{center}
\vspace{-20mm}
\epsfig{file=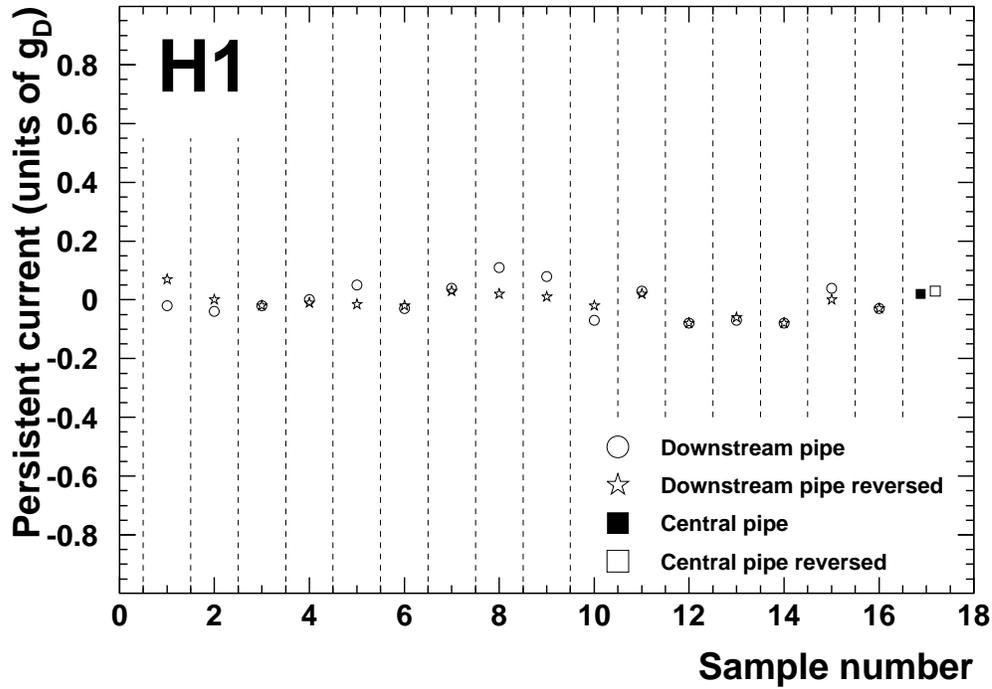,width=14.2cm}
\caption{The measured persistent currents in the long strips in units of 
$g_D$, after passage through the magnetometer, against sample 
number after the samples had been demagnetised (see text). Samples 1-16  
consisted of several long strips (usually two or three) 
from the downstream beam pipe bundled together. Sample 17 consisted of the 
thirteen long strips of the central beam pipe bundled together. These are 
shown individually in Fig. \ref{overview_central} before demagnetisation. 
All pieces of beam pipe tested are included in this 
plot except the short segments shown in Fig. \ref{overview_small}. 
None of the readings indicate the presence of a magnetic monopole. } 
\label{overview}
\end{center}
\end{figure}

\begin{figure}
\begin{center}
\vspace{-20mm}
\epsfig{file=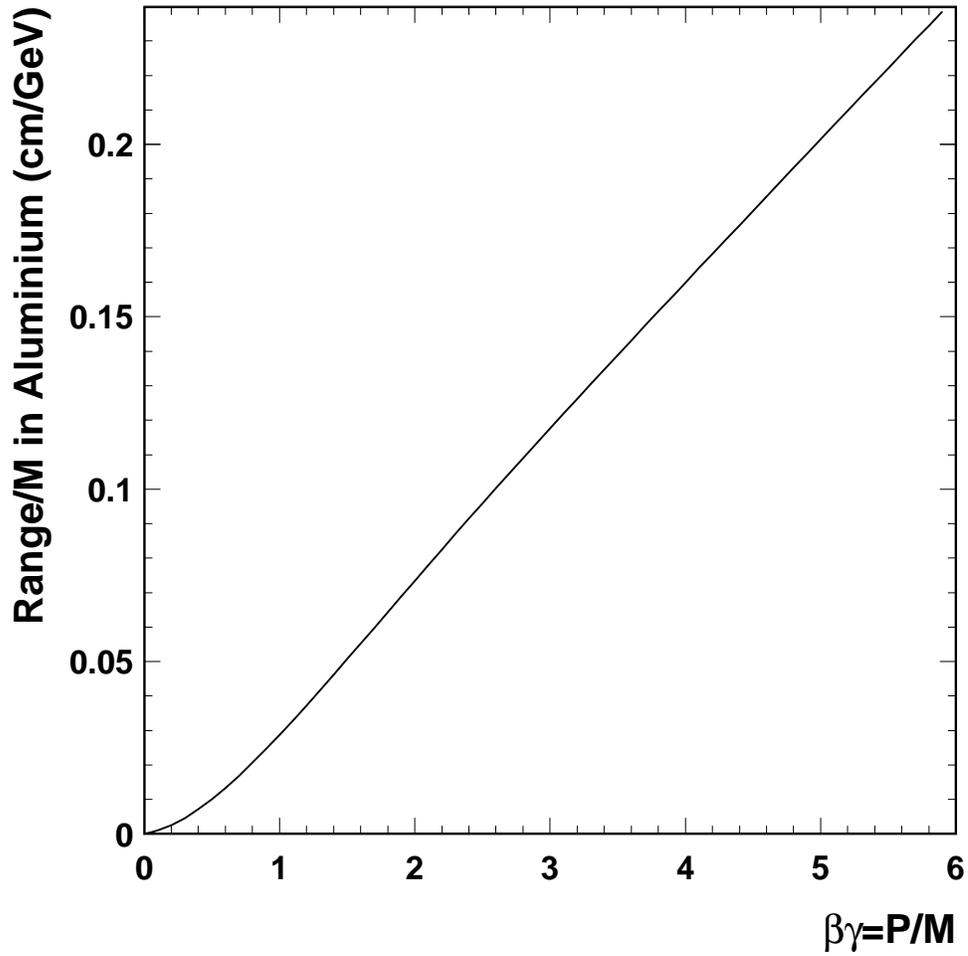,width=14.2cm}
\caption{The ratio of the range to mass of a monopole of charge $g_D$ 
in aluminium versus $\beta \gamma$. The range was calculated from the stopping 
power, $dE/dx$, in Fig. 1 of \cite{ahlen}, adjusted to the electron density 
in aluminium.}
\label{RErel}
\end{center}
\end{figure}

\begin{figure}[htb]
\begin{center}
\epsfig{file=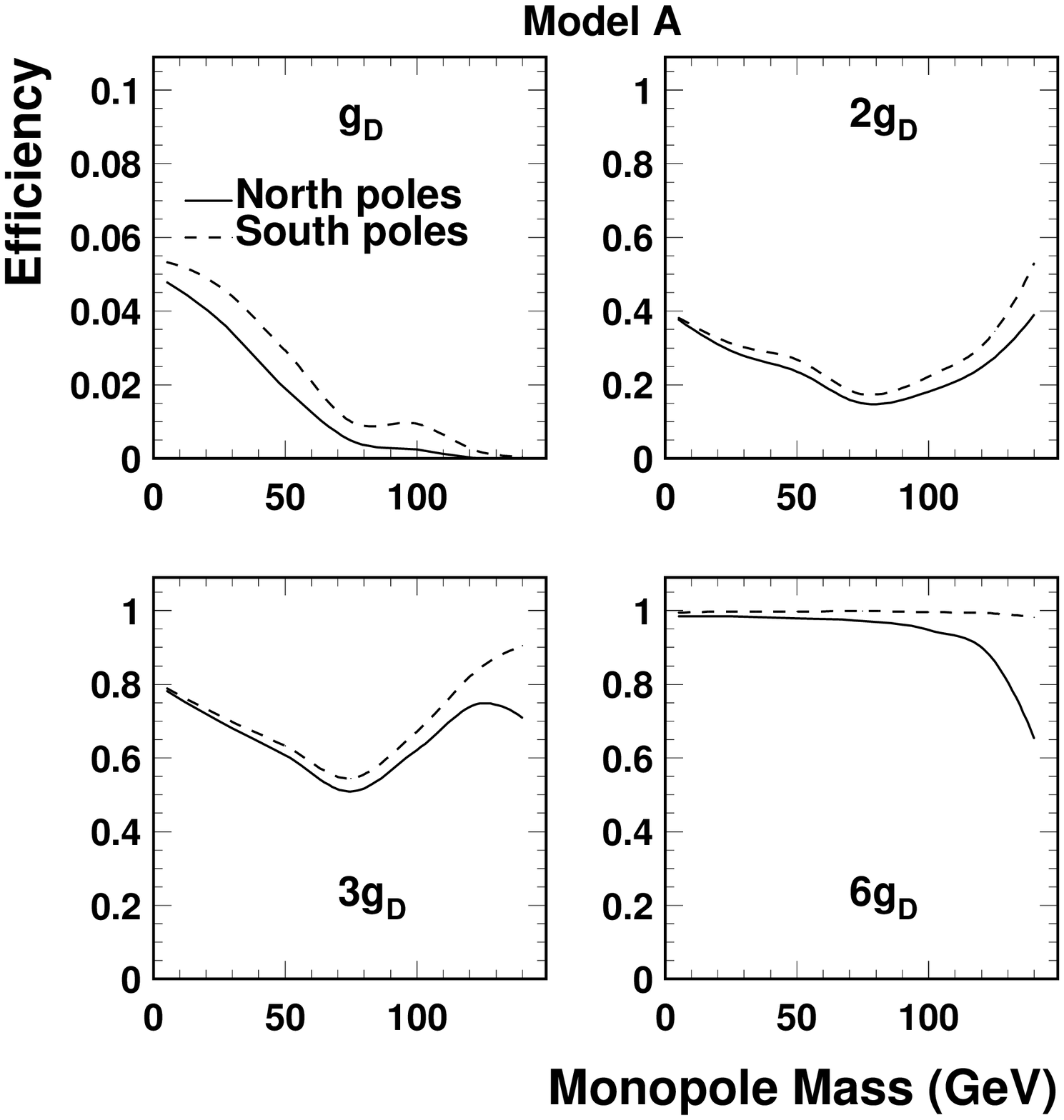,width=14.cm}
\caption{The efficiency for stopping monopoles of strength $g_D$,
$2g_D$, $3g_D$ and $6g_D$ or more computed according to the 
monopole pair production model A.}
\label{DISaccB}
\end{center}
\end{figure}

\begin{figure}[htb]
\begin{center}
\epsfig{file=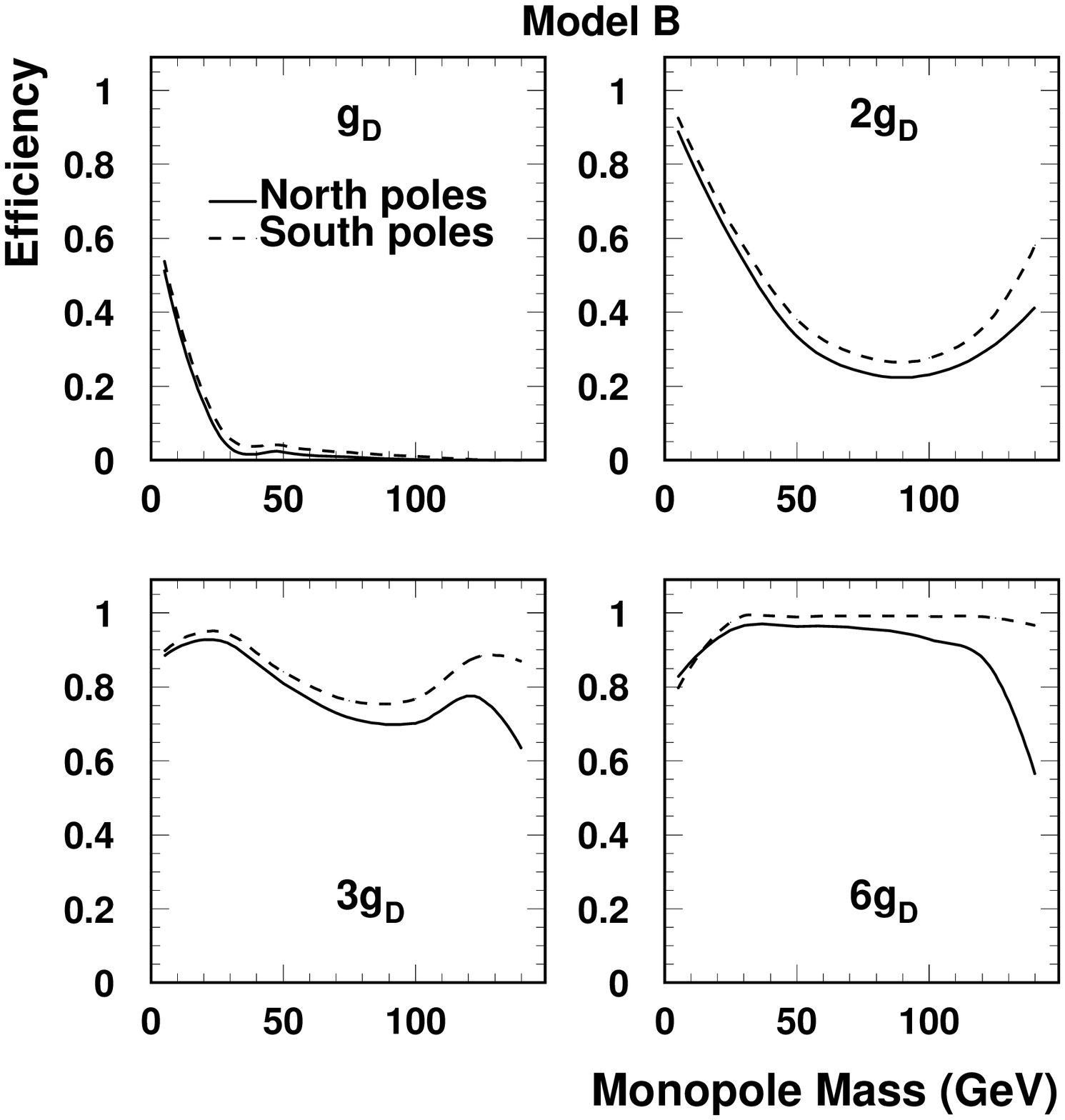,width=14.cm}
\caption{The efficiency for stopping monopoles of strength $g_D$,
$2g_D$, $3g_D$ and $6g_D$ or more computed according to the monopole pair 
production model B.}
\label{DISaccF}
\end{center}
\end{figure}

\begin{figure}[htb]
\begin{center}
\epsfig{file=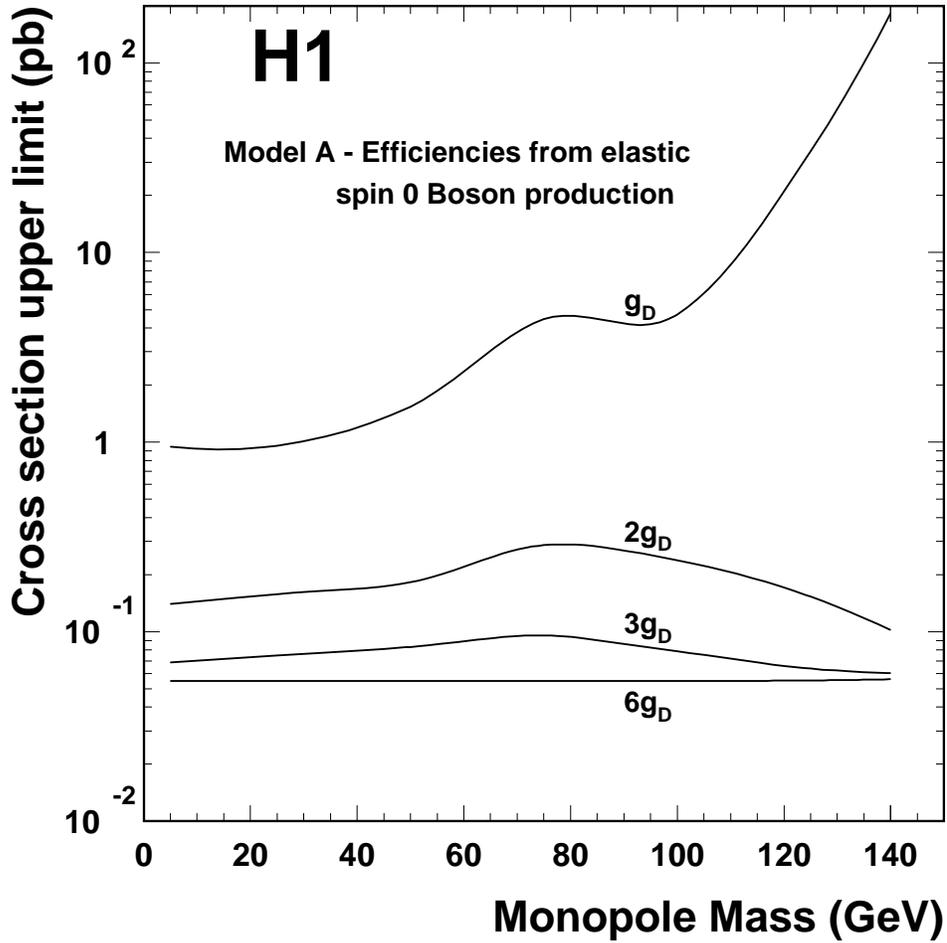,width=14.cm}
\caption{Upper limits on the cross section, determined within the context 
of model A, for monopole-antimonopole
pair production in $e^+p$ collisions as a function of monopole mass for 
monopoles of strength $g_D$, $2g_D$, $3g_D$ and $6g_D$ or more.}
\label{ltBoson}
\end{center}
\end{figure}

\begin{figure}[htb]
\begin{center}
\epsfig{file=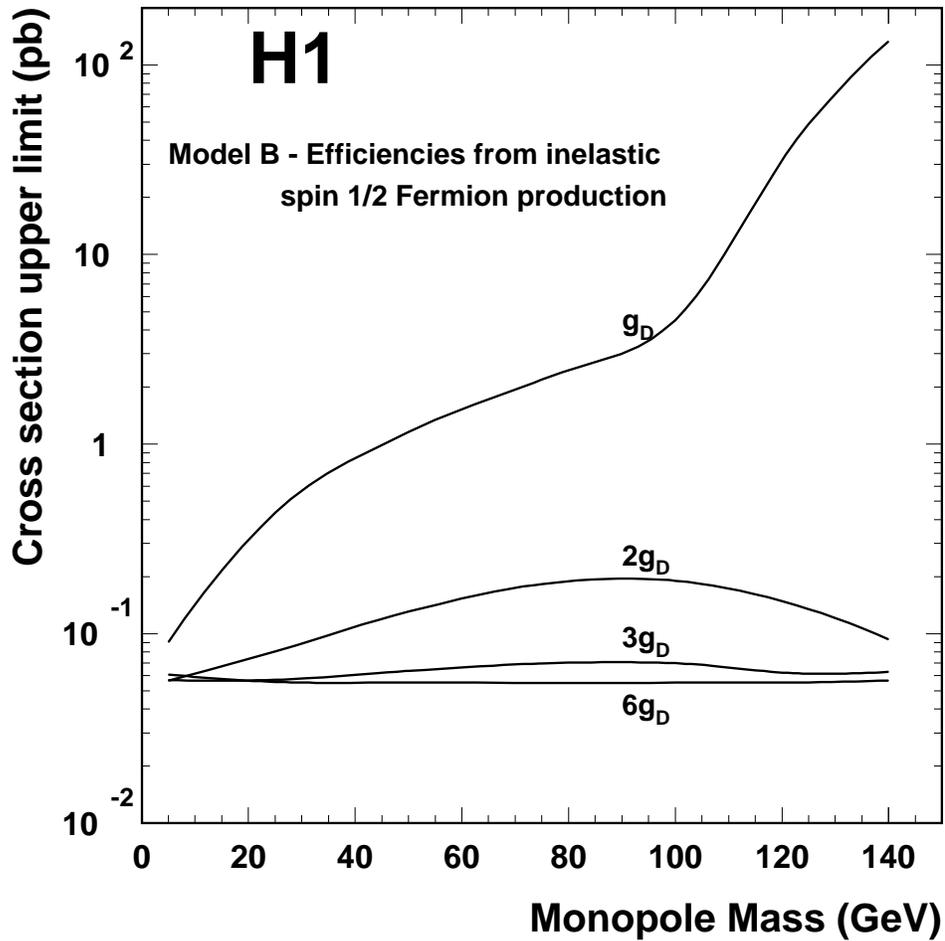,width=14.cm}
\caption{Upper limits on the cross section for monopole-antimonopole
pair production in $e^+p$ collisions, determined within the context 
of model B, as a function of monopole mass for 
monopoles of strength $g_D$, $2g_D$, $3g_D$ and $6g_D$ or more.}
\label{ltFermion}
\end{center}
\end{figure}